\documentclass[tightenlines,a4paper,fleqn,altaffilletter,twoside,twocolumn,
               floatfix,nofootinbib]{revtex4}

\usepackage[english]{babel}
\usepackage{amsmath,amssymb}
\usepackage{graphicx}
\usepackage{url}
\usepackage{epsfig}

\begin{document}

% =============================================================================
\title{The GSI oscillation mystery}
\author{Alexander Merle} \email[Email: ]{Alexander.Merle@mpi-hd.mpg.de}
\affiliation{Max--Planck--Institut f\"ur Kernphysik\\
             Postfach 10 39 80, D--69029 Heidelberg, Germany}
% =============================================================================

\begin{abstract}
In this talk, a short discussion of the GSI anomaly is given. We discuss the physics involved using a comparison with pion decay, and explain why the observed oscillations cannot be caused by standard neutrino mixing.\\

\end{abstract}

% -----------------------------------------------------------------------------
\maketitle
% -----------------------------------------------------------------------------

%%%%%%%%%%%%%%%%%%%%%%%%%%%%%%%%%%%%%%%%%%%%%%%%%%%%%%%%%%%%%%%%%%%%%%
\section{\label{sec:intro} What has been observed at GSI}
%%%%%%%%%%%%%%%%%%%%%%%%%%%%%%%%%%%%%%%%%%%%%%%%%%%%%%%%%%%%%%%%%%%%%%

This talk is based on Refs.~\cite{Merle:2009re,Kienert:2008nz}. In Ref.~\cite{Litvinov:2008rk} it has been reported that, in electron capture (EC) decays, several highly charged ions may show a decay law modified by superimposed oscillations. After this, a huge debate arose in literature as to whether this observation could be related to neutrino mixing, or not~\cite{Lipkin:2008ai,Ivanov:2008sd,Giunti:2008ex,Ivanov:2008nb,Faber:2008tu,Walker:2008zzb,Ivanov:2008xw,Kleinert:2008ps,Ivanov:2008zzc,Burkhardt:2008ek,Peshkin:2008vk,Giunti:2008im,Lipkin:2008in,Vetter:2008ne,Litvinov:2008hf,Ivanov:2008zn,Faestermann:2008jt,Giunti:2008eb,Gal:2008sw,Pavlichenkov:2008tm,Cohen:2008qb,Peshkin:2008qz,Lambiase:2008ki,Giunti:2008db,Lipkin:2009zy,Ivanov:2009rc,Faber:2009mg,Isakov:2009yr,Winckler:2009jm,Ivanov:2009en,Flambaum:2009di,Ivanov:2009kt,Ivanov:2009ku}.

%%%%%%%%%%%%%%%%%%%%%%%%%%%%%%%%%%%%%%%%%%%%%%%%%%%%%%%%%%%%%%%%%%%%%%
\section{\label{sec:supoerpos} The superposition principle}
%%%%%%%%%%%%%%%%%%%%%%%%%%%%%%%%%%%%%%%%%%%%%%%%%%%%%%%%%%%%%%%%%%%%%%
\begin{figure}[t]
\centering
  \epsfig{file=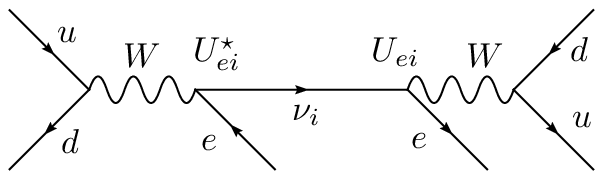}\ \ \ \
  \epsfig{file=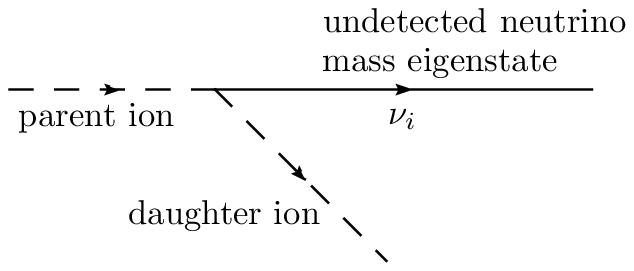}
  \caption{The Feynman diagrams for neutrino oscillations (a) and for EC (b).}
\end{figure}

The superposition principle states that, if several Feynman diagrams lead from the same initial to the same final state, the corresponding amplitudes have to be summed before squaring the total amplitude ({\it coherent summation}), while otherwise one has to sum over probabilities ({\it incoherent summation}). The former can lead to interference terms as; e.g., in neutrino oscillations, see Fig.~1(a). In the latter, the different final states are physically distinct and cannot interfere, as for the different neutrino mass eigenstates that can be emitted in an EC decay (see Fig.~1(b)).

%%%%%%%%%%%%%%%%%%%%%%%%%%%%%%%%%%%%%%%%%%%%%%%%%%%%%%%%%%%%%%%%%%%%%%
\section{\label{sec:amps} Probability Amplitudes}
%%%%%%%%%%%%%%%%%%%%%%%%%%%%%%%%%%%%%%%%%%%%%%%%%%%%%%%%%%%%%%%%%%%%%%

%%%%%%%%%%%%%%%%%%%%%%%%%%%%%%%%%%%%%%%%%%%%%%%%%%%%%%%%%%%%%%%%%%%%%%
\subsection{\label{sec:pion} First example: Charged pion decay}
%%%%%%%%%%%%%%%%%%%%%%%%%%%%%%%%%%%%%%%%%%%%%%%%%%%%%%%%%%%%%%%%%%%%%%

Let us go away from the GSI oscillations and consider the decay of a charged pion (e.g.\ $\pi^+$) into $|\mu^+ \nu_\mu \rangle$ or $|e^+ \nu_e\rangle$. An initial pion $|\pi^+\rangle$ evolves with time, and the corresponding time-dependent state $|\Psi (t)\rangle$ will be a coherent superposition of all parent and daughter states,
\begin{equation}
 |\Psi (t)\rangle=\mathcal{A}_\pi (t) |\pi^+\rangle+ \mathcal{A}_\mu (t) |\mu^+ \nu_\mu\rangle+ \mathcal{A}_e (t)|e^+ \nu_e\rangle,
 \label{eq:pion_1}
\end{equation}
whose correct normalization is imposed by the condition $|\mathcal{A}_\pi (t)|^2+|\mathcal{A}_\mu (t)|^2+|\mathcal{A}_e (t)|^2=1$. Note that the basis states are orthogonal, as they are clearly distinct. The measurement process projects this time-evolved state onto some measured state $|\Psi'\rangle$ (with a corresponding probability $P=|\langle \Psi'|\Psi (t) \rangle|^2$) that depends on the respective experiment. Let us discuss several cases:
\begin{itemize}

\item The (trivial) case is that there has been no detection at all: Then we have gained no information. This means that the projected state is just the time-evolved state itself, $|\Psi'\rangle=|\Psi (t)\rangle$, as the probability for anything to happen must be equal to 1.

\item If the experimental apparatus can give us the information that the pion has decayed, but one does not know the exact final state, then the only information is that $\mathcal{A}_\pi=0$ in Eq.~\eqref{eq:pion_1}. This leads to
\begin{equation}
 | \Psi'\rangle = \frac{\mathcal{A}_\mu (t) |\mu^+ \nu_\mu\rangle+ \mathcal{A}_e (t)|e^+ \nu_e\rangle}{\sqrt{|\mathcal{A}_\mu (t)|^2+|\mathcal{A}_e (t)|^2}}
 \label{eq:pion_2}
\end{equation}
with $P=|\mathcal{A}_\mu (t)|^2+|\mathcal{A}_e (t)|^2$. If there is any oscillatory phase in the amplitudes, $\mathcal{A}_k (t)= \mathcal{\tilde A}_k (t) e^{i\omega_k t}$, it will have no effect due to the absolute values.

\item If we know that the initial pion is still present, this simply sets $\mathcal{A}_\mu (t)=\mathcal{A}_e (t)=0$, and $|\Psi'\rangle=\mathcal{A}_\pi (t) |\pi^+\rangle /\sqrt{|\mathcal{A}_\pi (t)|^2}$ with $P=|\mathcal{A}_\pi (t)|^2$, which again does not oscillate.

\item If one particular final state, let us say $|e^+ \nu_e\rangle$, is detected, then $\mathcal{A}_\pi (t)=\mathcal{A}_\mu (t)=0$ and we get another term free of oscillations: $P=|\langle \Psi|\pi^+ (t) \rangle|^2=|\mathcal{A}_e (t)|^2$.

\end{itemize}

%%%%%%%%%%%%%%%%%%%%%%%%%%%%%%%%%%%%%%%%%%%%%%%%%%%%%%%%%%%%%%%%%%%%%%
\subsection{\label{sec:neutrino} Second example: Neutrinos}
%%%%%%%%%%%%%%%%%%%%%%%%%%%%%%%%%%%%%%%%%%%%%%%%%%%%%%%%%%%%%%%%%%%%%%

To be consistent with the GSI experiment, we consider a hydrogen-like ion as initial state $|M\rangle$ that can decay to the state $|D \nu_e\rangle$ via electron capture. Having factored out the leptonic mixing matrix elements $U_{ei}$, the time-evolution of the initial state is given by:
\begin{equation}
 |\Psi(t)\rangle = \mathcal{A}_M (t) |M\rangle + U_{e1} \mathcal{A}_1 (t) |D \nu_1\rangle+U_{e2} \mathcal{A}_2 (t) |D \nu_2\rangle
 \label{eq:neutrino_1}
\end{equation}
with $|\mathcal{A}_M (t)|^2+|U_{e1} \mathcal{A}_1 (t)|^2+|U_{e2} \mathcal{A}_2 (t)|^2=1$, which has exactly the same form as Eq.~\eqref{eq:pion_1}. We can immediately look at different cases:
\begin{itemize}

\item If the parent ion is seen in the experiment, this kills all daughter amplitudes, $\mathcal{A}_{1,2} (t)=0$. The only remaining amplitude is $\mathcal{A}_M (t)$, which leads to the non-oscillating probability $P=|\mathcal{A}_M (t)|^2$.

\item The next case corresponds to the GSI experiment: One measures the decay, but cannot tell which of the two neutrino mass eigenstates has been produced. This leads to $\mathcal{A}_M (t)=0$ and one has to perform a projection onto the state
\begin{equation}
 |\Psi' \rangle = \frac{U_{e1}\mathcal{A}_1 (t) |D \nu_1\rangle+U_{e2}\mathcal{A}_2 (t)|D \nu_2\rangle}{\sqrt{|U_{e1}\mathcal{A}_1 (t)|^2+|U_{e2}\mathcal{A}_2 (t)|^2}},
 \label{eq:neutrino_3}
\end{equation}
which yields $P=|U_{e1}\mathcal{A}_1 (t)|^2+|U_{e2}\mathcal{A}_2 (t)|^2$. Again, the result is an incoherent sum over probabilities, which exactly reflects the superposition principle. As in the probability corresponding to Eq.~\eqref{eq:pion_2} any oscillatory phase will drop out.

\item A hypothetical GSI-like experiment with infinite kinematical precision could distinguish the states $|D \nu_1\rangle$ and $|D \nu_2\rangle$. If one knows, e.g., that $|D \nu_1\rangle$ has been produced, one will have yet another term without oscillations, $P=|U_{e1} \mathcal{A}_1 (t)|^2$.

\end{itemize}

These are all cases that can appear. One can furthermore show that the neutrino that is emitted in the GSI experiment (of course) does oscillate, which is done in Ref.~\cite{Merle:2009re}, but this does not affect the lifetime of the initial ion in any way. The only known physical possibility that could actually explain an oscillatory behavior of the lifetime is a tiny energy splitting in the initial state~\cite{Kienert:2008nz,Giunti:2008ex,Peshkin:2008vk,Giunti:2008im}, whose origin has, however, not been explained yet.

%%%%%%%%%%%%%%%%%%%%%%%%%%%%%%%%%%%%%%%%%%%%%%%%%%%%%%%%%%%%%%%%%%%%%%
\section{\label{sec:conclusions} Conclusions}
%%%%%%%%%%%%%%%%%%%%%%%%%%%%%%%%%%%%%%%%%%%%%%%%%%%%%%%%%%%%%%%%%%%%%%

It has been shown that the oscillatory modulation of the decay law that has been observed in the GSI experiment cannot be explained by standard neutrino mixing. This has been done using probability amplitudes that turn out to be the most convenient language to use. Although there have been many attempts for an explanation of the GSI anomaly, a satisfactory one is still missing.

%%%%%%%%%%%%%%%%%%%%%%%%%%%%%%%%%%%%%%%%%%%%%%%%%%%%%%%%%%%%%%%%%%%%%%
\section*{\label{sec:ack} Acknowledgements}
%%%%%%%%%%%%%%%%%%%%%%%%%%%%%%%%%%%%%%%%%%%%%%%%%%%%%%%%%%%%%%%%%%%%%%

The author would like to thank his collaborators for the work done together, H.~Steiner for a useful comment on the notation, the organizers of the Erice 2009 meeting ``Neutrinos in Cosmology, in Astro, Particle and Nuclear Physics'' for doing an excellent job, as well as the participants of the workshop for many interesting and illuminating discussions. Furthermore, the author is grateful for having received a participation fellowship by DFG (Deutsche Forschungsgemeinschaft).

\end{document}